\documentclass[twocolumn]{aastex631}

\usepackage[normalem]{ulem}
\usepackage{color} 

\received{\today}
\graphicspath{{../figures+/}}

\begin{document}

\title{RNAAS \\
  Einasto gravitational potentials have difficulty to hold spherically \\
  symmetric stellar systems with cores
} 

\correspondingauthor{JSA}
\email{jos@iac.es}

\author[0000-0003-1123-6003]{Jorge S\'anchez Almeida} \affil{Instituto de Astrof\'\i sica de Canarias, La Laguna, Tenerife, E-38200, Spain} \affil{Departamento de Astrof\'\i sica, Universidad de La Laguna}



\begin{abstract}
It was known that an ideal spherically symmetric  stellar system with isotropic velocities and an inner core cannot reside in a Navarro, Frenk, and White (NFW) gravitational potential. The incompatibility can be pinned down to the radial gradient of the NFW potential in the very center of the system, which differs from zero. The gradient is identically zero in an Einasto potential, also an alternative representation of the dark matter (DM) halos created by the kind of cold DM (CDM) defining the current cosmological model. Here we show that, despite the inner gradient being zero, stellar cores are also inconsistent with Einasto potentials. This result may have implications to constrain the nature of DM through interpreting the stellar cores often observed in dwarf galaxies.   
\end{abstract}



\section{Discusion}\label{sec:intro}

\begin{figure*}[ht!] 
\centering
\includegraphics[width=0.8\linewidth]{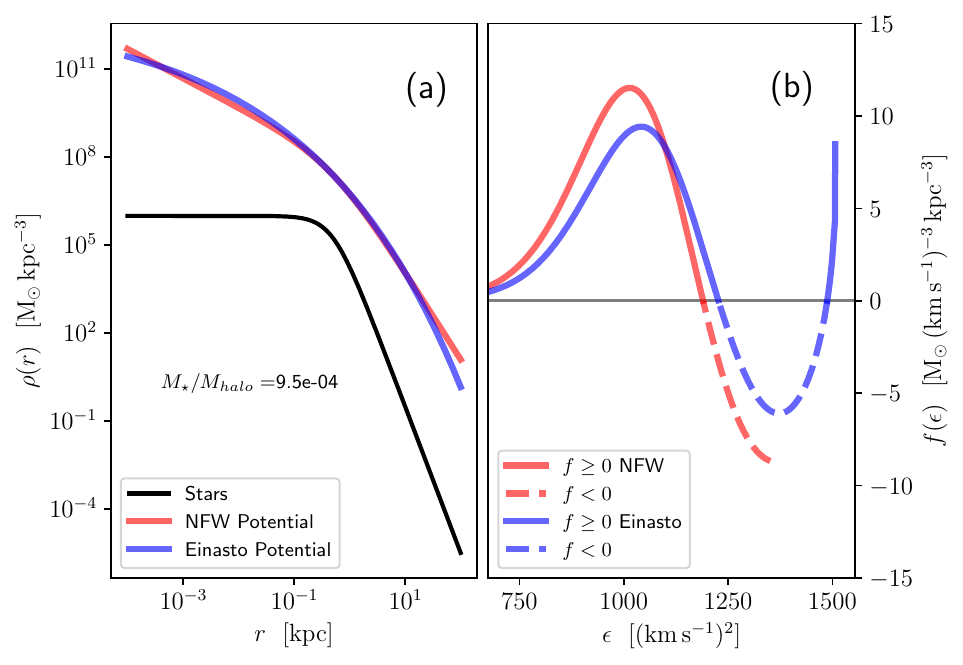}
\caption{
(a) Stellar density profile with a core (the black line) together with the mass density giving rise to a NFW potential (the red line) and an Einasto potential (the blue line). (b) Distribution function required from the stellar profile to reside in the corresponding NFW potential (the red line) and in the Einasto potential (the blue line). In both cases, $f(\epsilon)$ becomes negative  (the dashed lines).  The parameters were chosen to represent a dwarf galaxy with $M_\star\simeq 5\times 10^5\,M_\odot$.
}
\label{fig:fig1}
\end{figure*}
\citet{2023ApJ...954..153S} showed that an ideal spherical stellar distribution having an inner core, i.e., having
\begin{equation}
  \lim_{r\to 0}\frac{d\rho}{dr} = 0,
  \label{eq:core}
\end{equation}
is not consistent with a NFW potential \citep[named after][]{1997ApJ...490..493N}.  The symbol  $\rho$ stands for the stellar mass volume density, with $r$ the distance to the center of the gravitating system. This result would be only a curiosity if it were not for the fact that stellar cores are very common in dwarf galaxies \citep[e.g.,][]{2020ApJ...892...27M,2024ApJ...967...72R,2024A&A...681A..15M}. Even if it is an academic result for ideal systems, the existence of cores seems to suggest the gravitational potential to differ from the NFW potential characteristic of the CDM particles defining the current cosmological model. Deviations from the NFW potential are well known from kinematical measurements \citep[e.g.,][]{2015AJ....149..180O} and they are ascribed to the effect of stellar feedback processes modifying the overall gravitational potential \citep[e.g.,][]{2010Natur.463..203G}. This baryon driven feedback is expected to be inoperative for galaxies with  stellar mass $M_\star \ll 10^6\,M_\odot$ \citep[e.g.,][]{2015MNRAS.454.2981C,2021MNRAS.502.4262J}, therefore, the existence of stellar cores in tiny galaxies would reflect the need to go beyond the CDM paradigm. 

 \citet{2023ApJ...954..153S} proved the inconsistency using the Eddington Inversion Method \citep[EIM; e.g.,][]{2008gady.book.....B} which provides the distribution function in the phase space $f(\epsilon)$ needed to reproduce a stellar distribution in an assumed gravitational potential. For a cored stellar profile to reside in a NFW potential $f(\epsilon)$ has to be negative, which is unphysical. The proof employs an intermediate step in the EIM formalism, which reads,
\begin{equation}
  \frac{ d\rho/dr}{d\Psi/dr}=2\pi\sqrt{2}\,\int_0^{\Psi} \frac{f(\epsilon)}{\sqrt{\Psi-\epsilon}}\,d\epsilon,
  \label{eq:main}
\end{equation}
with $\Psi(r)$ the relative gravitational potential and $\epsilon$ the total energy per unit mass of a star. In the case of a NFW potential,
\begin{equation}
\lim_{r\to 0}\frac{d\Psi_{\rm NFW}}{dr} \not= 0,
\end{equation}
which together with Eq.~(\ref{eq:core}) implies the integral in the right-hand-side of Eq.~(\ref{eq:main}) to be zero  when $r\to 0$.
For the integral to be zero $f(\epsilon)$ must be negative somewhere in the phase space, which is unphysical and evidences the incompatibility. 

The CDM halos giving rise to a NFW potential are equally well represented by Einasto profiles \citep[e.g.,][]{2020Natur.585...39W}, where the derivative is zero \citep[e.g.,][]{2012A&A...540A..70R}
\begin{equation}
\lim_{r\to 0}\frac{d\Psi_{\rm Einasto}}{dr} = 0
\end{equation}
and, consequently, for which the above argument of incompatibility does not hold. 
The question arises as to whether the incompatibility of stellar cores is specific to the NFW potential or if it more generally affects all CDM-produced halos. Recall that the NFW potential stems from a mass density profile scaling as $r^{-1}$, therefore, with an artificial singularity  when $r \to 0$. The singularity is not present in the Einasto profile which  scales as $\exp\left[\left(r/h\right)^\alpha\right]$ with $\alpha$ and $h$ two positive constants. This {\em Note} is meant to show that the inconsistency still remains in the case of an Einasto potential. In terms of Eq.~(\ref{eq:main}), the Einasto profile has  
\begin{equation}
  \frac{d\Psi_{\rm Einasto}}{dr} \not= 0,
  \label{eq:core3}
\end{equation}
for values of $r\not=0$ where
\begin{equation}
  \frac{d\rho}{dr} \simeq 0,
  \label{eq:core2}
\end{equation}
and this condition is enough to force the need of $f(\epsilon) < 0$.

The argument is illustrated with the simulation in Fig.~\ref{fig:fig1}, which shows
a cored stellar profile (the black line; a Plummer profile) with size and central density chosen to mimic a dwarf galaxy. 
The red line shows the NFW profile used to compute the potential giving rise to the distribution function shown in Fig.~\ref{fig:fig1}b (the red line). It was evaluated  with the numerical tools developed by \citet{2023ApJ...954..153S}. As expected, there is a range of values for which $f(\epsilon)<0$.  The same exercise is carried out with an Einasto profile having the same characteristic radius and density as the NFW profile,  
as in CDM simulations  \citep[e.g.,][]{2020Natur.585...39W}. Firstly, we note that the corresponding $f(\epsilon)$ also has negative parts (the blue dashed line in Fig.~\ref{fig:fig1}b), making also the Einasto potential inconsistent with the stellar core. Secondly, and in contrast with the NFW potential, the distribution function for the Einasto potential becomes positive near the largest energy. Such energies are reached only by the stars moving close to the center of the potential at all times. Thus, the incompatibility is not at $r\to 0$ but at a finite radius  within the core (Eq.~[\ref{eq:core2}]) but where the potential keeps changing (Eq.~[\ref{eq:core3}]).

%
\section{Conclusions}

The original inconsistency happens for spherically symmetric systems having isotropic velocities, and representing the CDM with NFW profiles. However, none of these simplifying assumptions seem to be responsible for the incompatibility.  \citet{2023ApJ...954..153S} showed that the incompatibility remains for other potentials with an inner slope much shallower than the value of $-1$ characteristic of the NFW profiles. They also showed that the assumption of isotropic velocities can be relaxed since the incompatibility remains for radial orbits and for isotropic orbits that turn into radial in the outskirts, as expected from cosmological numerical simulations \citep[e.g.,][]{2017ApJ...835..193E,2023MNRAS.525.3516O}. On the other hand, \citet{sanchez_almeida_carlstein} proved that the incompatibility persists for axi-symmetric systems, which implies that it goes beyond the assumption of spherical symmetry.
This {\em Note} adds that the incompatibility remains for Einasto potentials, which are also a good representation of the  CDM halos and, in contrast to the NFW profiles, do not diverge at the center of the system. Thus, the incompatibility is not artificially set by the mathematical singularity at $r=0$  present in the NFW profiles.  

\begin{acknowledgments}
Thanks are due to Ignacio Trujillo and Angel Plastino for many insightful discussions on how the properties of the gravitational potential are constrained by the observed starlight. 
The Spanish Ministry of Science and Innovation supported the work through project PID2022-136598NB-C31 (ESTALLIDOS).
\end{acknowledgments}

%


\software{ {\tt Scipy} \citep{2020SciPy-NMeth}}





\end{document}